# Efficient implementation of immersed boundary-lattice Boltzmann method for massive particle-laden flows Part I: Serial computing

Maoqiang Jiang, Jing Li, Zhaohui Liu†

*State Key Laboratory of Coal Combustion, School of Energy and Power Engineering,*

*Huazhong University of Science and Technology, Wuhan 430074, PR China*

† Corresponding author. Tel.: +86-027-8754-2417-8308. E-mail address: zliu@hust.edu.cn (Z. Liu).

**Highlights:**

1. Symmetry algorithm is proposed for the half-calculation of LB collision and external force term.
2. Swap algorithm is used for the half-memory of LB PDFs.
3. Local point-to-node algorithm is proposed for the IBM calculation.
4. Improved grid search algorithm is proposed for searching particle-pair collisions.
5. High performances of present algorithms are obtained in 2D and 3D tests.

**Abstract:**

Immersed boundary-lattice Boltzmann method (IB-LBM) has been widely used for simulation of particle-laden flows recently. However, it was limited to small-scale simulations with no more than $O(10^3)$ particles. Here, we expand IB-LBM for massive particle-laden flows with more than $O(10^4)$ particles by two sequential works. First is the Part I: serial computing on a single CPU core and following the Part II: parallel computing on many CPU cores. In this Part I paper, a highly efficient and localized implementation of IB-LBM is proposed for serial computing. We optimize in three main aspects: swap algorithm for incompressible LBM, local grid-to-point algorithm for IBM and improved grid search algorithm for particle pair short-range interaction. In addition, symmetry algorithm is proposed for the half-calculation of LB collision and external force term. The computational performance on a single CPU core is analyzed. Different scales of two dimensional (2D) and three-dimensional (3D) particles settling in closed cavities are used for testing. The solid volume fraction is varied from 0 to 0.40. Simulation results demonstrate that all calculation parts are dramatically decreased by the improved algorithm. For the particle-free flows, the Mega Lattice Site Update per Second (MLUPS) can be achieved up to 36 (2D) and 12 (3D) using the improved algorithm. For the particle-laden flows, MLUPS can be achieved no lower than 15 (2D) and 7 (3D) in the simulations of dense flows. At last, we discuss the potential of the new algorithms for the high-performance computation of the large-scale systems of particle-laden flows with MPI parallel technique.

**Keywords:**

Direct numerical simulation; Boundary thickening based direct forcing; Immersed boundary method; Lattice Boltzmann method; Particle-laden flows

# 1. Introduction

Particle-laden flows in which fluid carries large number of dispersed particles, play a key role in various natural and industrial processes. On one hand, nature examples of such flows include cell motion in blood, sand and dust storms, volcanic eruptions, sand sedimentation and transport in river. On the other hand, engineering examples consist of particle suspended in fluidized beds, coal combustion in boilers, mineral core smelting in blast furnaces, pneumatic conveying of dust, etc. The multiscale and multi-physics characteristics involved make it complex and complicated to be understood. Fully (or particle) resolved direct numerical simulation (F/PR-DNS) based on immersed boundary-lattice Boltzmann method (IB-LBM), as one of the popular first-principle approaches for revealing the dynamic fundamentals, has rapidly become a focus in recent years [1, 2]. However, it is unrealistic for large scale simulations due to the high demands of memory and computing time, though supercomputer has been drastically developed. At present, most works focused on the accuracy improvement and physical field expansion development of this algorithm. A few literatures [3-6] had applied IB-LBM to study the real isothermal and non- isothermal particulate flows, while the number of particles are limit to no more than $O(10^3)$. It is the key to improve or optimize this specific method for application to achieve three goals: higher computational efficiency, lower memory footprint and flexibility for parallelization. Before putting effort into the parallelization of a code, the serial performance of the implementation of IB-LBM on a single processor should be optimized [7], which is the aim of this Part I paper.

Lattice Boltzmann method (LBM) has been successfully, effectively and widely applied to solve various complex flows over the past three decades [8, 9]. It was firstly adopted by Anthony Ladd [10, 11] in 1994 for the simulation of particle-laden flows. The main assets of LBM include simplicity of coding, straightforward incorporation of microscopic interactions, and suitability for parallel computing, while still at the expense of high computing cost, both in memory and computing time. Here, the most commonly used single relaxation time-lattice Bhatnagar-Gross-Krook (SRT-LBGK) scheme is used. In general, LBM is implemented by explicit time-marching on a uniform lattice with two steps (two-step algorithm): first collision on the local node and then streaming (or propagation) to the neighborhood nodes, in which the collision step is the most time-consuming and the streaming step is the most memory-consuming. Wellein [12] optimized the data layout for achieving high performance. However, it was not general because different optimization data layout should be adopted for different computer architectures. Martys & Hagedorn [13] proposed a simplified scheme by setting the relaxation parameter $\tau = 1$ that the storing of discrete particle distribution function (PDF) is not needed. However, it is non-universal because it is difficult to ensure $\tau = 1$ in all complex flows. Then, several works had focused on optimizing LBM computations especially for reducing the memory consumption [14], such as the compressed grid (shift) algorithm [15], A-A pattern [16], Esoteric twist algorithm [17] and swap algorithm [18]. The detail comparison of different schemes of streaming steps can be seen in Wittmann et al. [14, 19]. The swap algorithm was presented by Mattila et al.[18] and simultaneously used in the open source software OpenLB [20]. It implemented the D$d$Q$q$ model with only $q$ variables per node instead of $2q$, which can reduce the memory of the fluid particle distribution functions (PDFs) to be half. The streaming step is done by explicitly exchanging the PDFs values in opposite directions firstly at local nodes and then with that at neighboring nodes.

Immersed boundary method (IBM) [21, 22] has become the most popular method for FR-DNS of particle-laden flows over the last few years [23], especially with the large number of particles. It was firstly proposed by Peskin [24] in 1970s for simulation of blood flow with flexible valves inside the heart. It uses a fixed orthogonal Cartesian grid, while doesn't need a complicated unstructured grid for a complex geometry. And the computation intensive re-meshing process for moving geometries at each time step, which is commonly required in the arbitrary Lagrangian-Eulerian (ALE) approach, is also removed. The fluid is solved both outside and inside the particles, while enforcing a direct fictitious force on the local fluid nearby boundary to achieve non-slip conditions on each particle surface. Many works had used this method coupling with solver of Navier-Stokes equation (NSE) [25, 26] and LBM [27-30] for simulation of particle-laden flows. In those works, the proposed conventional scheme of the IBM is the direct forcing (DF) scheme [25, 27], which is simple and easy to be implemented but the non-slip boundary conditions on the particle surface cannot be satisfied. Subsequently, a multi-direct forcing (MDF) scheme [26, 28] was proposed to improve the non-slip boundary conditions but with a high computing cost due to the multi iteration of the IBM calculation. Recently, Jiang and Liu [31] proposed a high accuracy and efficient IBM scheme with low computing cost, called the boundary-thickening based direct forcing (BTDF) scheme. Though IBM has high efficiency for mesh generation compare to the ALE approach, the additional IBM computing on the particle boundaries is time consuming, especially for the high Reynolds number flows. Therefore, it is important to reduce the amount of this additional IBM computing. Unfortunately, there are very limited literatures reporting the detail implementation of the IBM computing, especially for the particle-laden flows。

For the flows laden with large number of particles, one additional problem should be considered: the short-range interactions between pair particles when they approaching. These short-range interactions are generally considered firstly by the lubrication force when the fluid between pair particles cannot be resolved, and further by the dry collision force when the pair particles directly collide[32]. In a large system with large number of particles, the tracking of these interactions is time consuming. For example, for a set of $N$ particles, the intrinsically searching complexity of the potential interaction pairs is ordered as $O(N^2)$ in each time step. When $N$ is more than $O(10^4)$, the computing task become extremely heavy. In fact, one given particle can only shortly interacting with no more than six equal-size surrounding particles, which can avoid large unnecessary computations. Based on this, many fast searching algorithms had been developed for the direct dry collisions in the particle simulations, such as those in macro-scale Discrete element method (DEM) and micro-scale molecular dynamic (MD) in the last century [33]. The well-known algorithm is the boxing method [34], in which the computational domain is divided into two or three dimensional small boxes, so-called "cells". They manage the discrete position of each particle. Since the search region for each particle is narrowed to cells in the vicinity of the target particle, this method reduces the complexity to $O(N)$. The boxing algorithms can be further categorized as the linked-cell algorithm and the grid search algorithm[33-35]. In the former [36-38], the searching cells are slightly or moderately larger than the particle size and so each of these cells is assigned a list of particles which reside in the cell. The calculation contains two steps: first the calculation of the interaction with other particles located in the same cell and then the interaction with the particles located in the neighboring cells. In the latter [39-41], the searching cells are smaller than the particle size to require that no cell is occupied by the center of more than one particle. Each cell is assigned a number which is equal to the index of the particle whose center resides at this cell, or -1 if there is no such particle. The calculation can only be conducted the interaction with the particles located in the neighboring cells. Compare with the former linked-cell algorithm, the grid search algorithm is simpler.

As far as we know, there are few literatures reporting the detail implementation and optimization of the coupled immersed boundary-lattice Boltzmann method (IB-LBM) for high efficient computing of particle-laden flows. In this paper, the detail implementation and according improvement in above three parts: LBM, IBM and particle collisions are conducted. In the first part, the swap algorithm is firstly used for half memory consumption because it is general. All the data is always preserved for parallelization and no data can be lost on the boundary. Furthermore, unrolling technique is presented for reducing calculation of macro quantity and LB collision. And the symmetrical characteristic of the discrete lattice velocity in the opposite directions is utilized to half the computing of the equilibrium PDFs and the external force terms in collision step. In the second part, a local point-to-node algorithm is proposed to implement an extremely simple computing of IBM. In the last part, the grid search algorithm is adopted and improved to further consider the lubrication interaction of approaching pair particles. It should be said that the algorithm optimizations reported here are not designed for any particular processor or system architecture. Instead, the optimizations are generally applicable to modern processors, and aim at reduce the calculation amount and improve the calculation efficiency.

This paper is organized as follows. Section 2 presents the numerical simulation model of the boundary-thickening based direct forcing IB-LBM, and the control equation for particle motion and particle-particle interaction. Section 3 to 5 introduced the optimization of the three aspects of detail implementation, respectively. The performance results and discussions are presented in Section 6. Finally, the conclusions and a further discussion are given in Section 7.

## 2. Description of the numerical method

The numerical method for fully resolved simulation of flows laden with finite size particles consists three parts: LBM for the fluid flow, IBM for the particle-fluid interaction, direct solution for particle movement and particle pair short-range interaction. Followed sub-sections simply introduce these three parts.

### 2.1. Incompressible LBM for fluid flow

In this paper, the fluid flow is simulated by the SRT LBGK model proposed by Guo et al.[42] with external force term as

$$f_\alpha \left( \mathbf{x} + \mathbf{e}_\alpha \Delta t, t + \Delta t \right) = f_\alpha \left( \mathbf{x}, t \right) - \frac{1}{\tau} \left[ f_\alpha \left( \mathbf{x}, t \right) - f_\alpha^{(eq)} \left( \mathbf{x}, t \right) \right] + F_\alpha \left( \mathbf{x}, t \right) \Delta t \tag{1}$$

where $f_\alpha(\mathbf{x}, t)$ is the particle distribution function (PDF) of the discrete velocity $\mathbf{e}_\alpha$ in the $\alpha$-th direction. $\Delta t$ is the time step and $\tau$ is the dimensionless relaxation time determined by fluid kinematic viscosity as $\tau = 0.5 + 3\upsilon \Delta x^2 / \Delta t$. In this paper, frequently used D2Q9 and D3Q19 model are adopted for the two dimensional and three-dimensional simulations respectively, as shown in Fig. 1. Its general formula is D$d$Q$q$, in which $d$ denotes the dimension and $q$ denotes the number of the discrete velocity directions. The discrete velocity vectors of 2D and 3D flows are defined as

$$[\mathbf{e}_{\alpha=0\text{-}8}] = c \begin{bmatrix} 0 & 1 & 1 & 0 & -1 & -1 & -1 & 0 & 1 \\ 0 & 0 & 1 & 1 & 1 & 0 & -1 & -1 & -1 \end{bmatrix} \quad (2)$$

and

$$[\mathbf{e}_{\alpha=0\text{-}18}] = c \begin{bmatrix} 0 & 1 & 0 & 0 & 1 & -1 & 0 & 0 & 1 & -1 & -1 & 0 & 0 & -1 & 1 & 0 & 0 & -1 & 1 \\ 0 & 0 & 1 & 0 & 1 & 1 & 1 & -1 & 0 & 0 & 0 & -1 & 0 & -1 & -1 & -1 & 1 & 0 & 0 \\ 0 & 0 & 0 & 1 & 0 & 0 & 1 & 1 & 1 & 0 & 0 & -1 & 0 & 0 & -1 & -1 & -1 & -1 \end{bmatrix} \quad (3)$$

where the lattice speed $c = \Delta x / \Delta t$, and $\Delta x$ is the lattice spacing step. The law of the arrangement of the discrete velocity vectors is that the $\mathbf{e}_\alpha = -\mathbf{e}_{\alpha-(q-1)/2}$ when $(q-1)/2 < \alpha < q$, which can be useful for the code optimization in Section 3. Woodgate et al. [43] shows that there is no difference on the simulation results for the different ordering of the lattice vectors.

In this incompressible LBM model, the equilibrium distribution function $f_\alpha^{(eq)}$ is defined as

$$f_\alpha^{(eq)} = \lambda_\alpha p + \rho_0 \omega_\alpha \left[ \frac{(\mathbf{e}_\alpha \cdot \mathbf{u})}{c_s^2} + \frac{(\mathbf{e}_\alpha \cdot \mathbf{u})^2}{2c_s^4} - \frac{|\mathbf{u}|^2}{2c_s^2} \right] \quad (4)$$

where $\lambda_\alpha$ is the model parameter satisfying $\lambda_0 = (\omega_0 - 1)/c_s^2$, $\lambda_\alpha = \omega_\alpha / c_s^2$ $(\alpha > 0)$ with the constant $\rho_0$ being the fluid density. The weighting coefficients $w_\alpha$ is $w_0 = 4/9$, $w_{1,3,5,7} = 1/9$, and $w_{2,4,6,8} = 1/36$ for 2D flows, and $w_0 = 1/3$, $w_{1\text{-}3,10\text{-}12} = 1/18$, and $w_{4\text{-}9,13\text{-}18} = 1/36$ for 3D flows respectively. The macro velocity $\mathbf{u}^*$ and pressure $p$ of fluid without external volume force are calculated respectively by

$$\mathbf{u}^* = \sum_\alpha \mathbf{e}_\alpha f_\alpha \quad (5)$$

and

$$p = \frac{c_s^2}{1 - \omega_0} \left( \sum_{\alpha \geq 1} f_\alpha - \rho_0 \omega_0 \frac{|\mathbf{u}|^2}{2c_s^2} \right) \quad (6)$$

When an external volume force term $\mathbf{f}(\mathbf{x},t)$ is introduced, we adopt the split forcing scheme from the Guo et al. [44] model by considering the discrete lattice effect. It corrects the velocity as

$$\mathbf{u} = \mathbf{u}^* + \frac{\Delta t}{2\rho} \mathbf{f}(\mathbf{x},t) \quad (7)$$

and corrects the distribution functions by the last term of the right hand in equation (1) as

$$F_\alpha(\mathbf{x},t) = \left(1 - \frac{1}{2\tau}\right) \omega_\alpha \left[ \frac{\mathbf{e}_\alpha - \mathbf{u}(\mathbf{x},t)}{c_s^2} + \frac{\mathbf{e}_\alpha \cdot \mathbf{u}(\mathbf{x},t)}{c_s^4} \mathbf{e}_\alpha \right] \cdot \mathbf{f}(\mathbf{x},t) \quad (8)$$

From equation (1) - (8), the incompressible governing equation can be correctly recovered by a Chapman–Enskog multi-scale expansion with second-order accuracy.

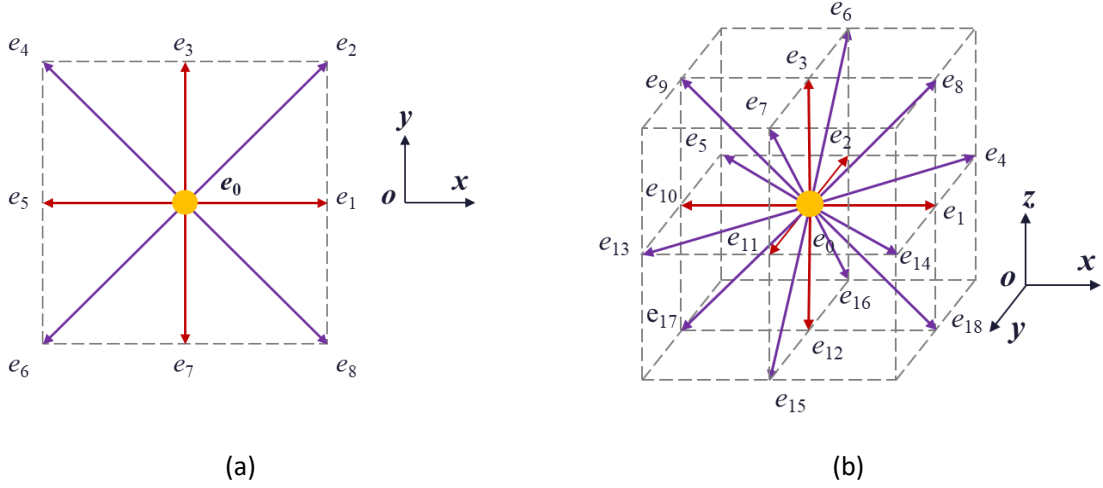

Fig. 1. The discrete lattice velocity vectors in (a) D2Q9 model and (b) D3Q19 model.

*2.2. BTDF-IBM for particle-fluid interaction*

BTDF-IBM is a simplified high efficient, low memory method proposed by Jiang & Liu [31] for getting satisfied non-slip and non-penetration conditions on particle boundaries. It numerically distinguishes the boundary to two overlapped shells with finite thickness: fluid boundary shell and solid boundary shell, as can be seen in Fig. 2(a). The thickness *drf* of the fluid boundary shell and the thickness *drs* of the solid boundary shell are decided by the selected regularized delta function in Fig. 2(b). In the conventional direct forcing (DF)-IBM, *drs* equals to one lattice step *dh* by default. This cause the non-conservation problem of the immersed boundary force between the fluid domain and the solid boundary, further causes the un-satisfaction in non-slip and non-penetration boundary conditions. In BTDF scheme, Jiang & Liu [31] indicated that the thickness of the solid boundary shell should be larger than one lattice step. They deduce different solid boundary thickness according to different regularized delta functions based on the conservation condition of the immersed boundary force. Here we simply introduce it while the details can be found in that paper.

BTDF-IBM mainly contains three steps: non-forcing velocity interpolation from fluid nodes to boundary points, IB force calculation on boundary points and IB force spreading from boundary points to fluid nodes, which can be expressed by Equations (9) to (11), respectively.

$$\mathbf{u}_b^* = D_I \mathbf{u}^* \tag{9}$$

$$\mathbf{F}_b = \frac{2\rho_f}{\Delta t}\left(\mathbf{U}_b - \mathbf{u}_b^*\right) \tag{10}$$

$$\mathbf{f} = D_E \mathbf{F}_b \tag{11}$$

where $\mathbf{U}_b$ and $\mathbf{F}_b$ are boundary velocity and IB force, respectively. $\mathbf{u}_b^*$ is the predicted velocity on boundary and $\mathbf{f}$ is the IB force on neighborhood fluid domain. $D_I$ is the interpolation operator matrix from fluid Eulerian nodes to Lagrangian points on solid boundary with a dimension of $N_L \times N_E$, and $D_E$ is the spreading operator matrix from Lagrangian points to Eulerian nodes with a dimension of $N_E \times N_L$. $N_E$ and $N_L$ are the total number fluid nodes and the number of Lagrangian points on particle surface, respectively.

The matrix elements of $D_I$ and $D_E$ are constructed by

$$D_{I,ij} = \frac{1}{dh^2} \delta_{ij}\left(\frac{x_i - X_j}{dh}\right) \delta_{ij}\left(\frac{y_i - Y_j}{dh}\right) \cdot dh^2 \tag{12}$$

$$D_{E,ji} = \frac{1}{dh^2} \delta_{ij}\left(\frac{x_i - X_j}{dh}\right) \delta_{ij}\left(\frac{y_i - Y_j}{dh}\right) \cdot (drs \cdot ds) \tag{13}$$

where $i$ varies from 1 to $N_L$, and $j$ varies from 1 to $N_E$, respectively. $\delta_{ij}(\cdot)$ is the regularized delta function. $ds$ is the distance between two adjacent Lagrangian points. $dh$ and $drs$ are the fluid lattice step and the solid boundary thickness. It should be pointed that in the conventional DF-IBM, boundary thickness $drs/dh$ equals to one. However, in the improved BTDF-IBM, the value of $drs/dh$ is larger than one corresponding to the different delta functions. For example, in general $drs/dh$ =1.9 and 2.6 for the 3-point function and 4-point function, respectively. Hence, the term ($drs \cdot ds$) depicts the local annular area of the solid boundary shell controlled by a target Lagrangian point.

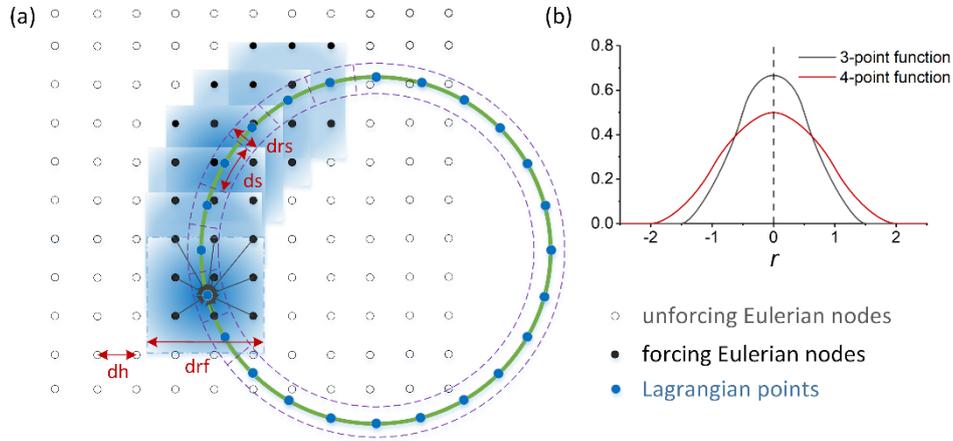

Fig. 2. The schematic illustration of (a) the BTDF immersed boundary method and (b) the two frequently used regularized delta functions.

*2.3. Particle movement and particle-pair short-range interaction*

The particles move not only with translational motion but also with rotational motion, which are controlled by the followed differential equations based on Newton's second law as

$$M_p \frac{d\mathbf{U}_c}{dt} = \mathbf{F}_h + \mathbf{F}_b + \mathbf{F}_c + \mathbf{F}_f \tag{14}$$

$$I_p \frac{d\mathbf{\Omega}_c}{dt} = \mathbf{T}_h + \mathbf{T}_f \tag{15}$$

in which $\mathbf{U}_c$ and $M_p$ are the translational velocity of particle center and the mass of the particle. $\mathbf{\Omega}_c$ and $I_p$ are the angular velocity and inertial moment of the particle. The first term of the right hand of Eq.(14) and Eq. (15) $\mathbf{F}_h$ and $\mathbf{T}_h$ are the hydrodynamic force and torque acted on the particles, which can be calculated by above BTDF-IBM [31]. $\mathbf{F}_f$ and $\mathbf{T}_f$ are the compensate virtual force to keep the interior fluid for rigid moving. $\mathbf{F}_b$ is the buoyant force and $\mathbf{F}_c$ is the particle-pair collision force. Here, we use the short-range repulsive force [45, 46] depicting this particle-pair (force on the $i$-th particle by the $j$-th particle) interaction as

$$\mathbf{F}_{ij}^{p} = \begin{cases} 0, & \|\mathbf{X}_{ci} - \mathbf{X}_{cj}\| \geq D_p + \zeta, \\ \dfrac{c_{ij}}{\varepsilon_{pp}}\left(\dfrac{\|\mathbf{X}_{ci} - \mathbf{X}_{cj}\| - D_p - \zeta}{\zeta}\right)^2 \left(\dfrac{\mathbf{X}_{ci} - \mathbf{X}_{cj}}{\|\mathbf{X}_{ci} - \mathbf{X}_{cj}\|}\right), & D_p \leq \|\mathbf{X}_{ci} - \mathbf{X}_{cj}\| < D_p + \zeta, \\ \left(\dfrac{c_{ij}}{\varepsilon_{pp}}\left(\dfrac{\|\mathbf{X}_{ci} - \mathbf{X}_{cj}\| - D_p - \zeta}{\zeta}\right)^2 + \dfrac{c_{ij}}{E_{pp}}\left(\dfrac{D_p - \|\mathbf{X}_{ci} - \mathbf{X}_{cj}\|}{\zeta}\right)\right)\left(\dfrac{\mathbf{X}_{ci} - \mathbf{X}_{cj}}{\|\mathbf{X}_{ci} - \mathbf{X}_{cj}\|}\right), & \|\mathbf{X}_{ci} - \mathbf{X}_{cj}\| < D_p \end{cases} \quad (16)$$

in which the parameter $\zeta$ denotes the threshold or the range of the repulsive force. $c_{ij}$ is the small-scale force which is set as the buoyancy force on the body. $\varepsilon_{pp}$ is the stiffness parameters and $E_{pp}$ is a smaller parameter than $\varepsilon_{pp}$ to ensure a much larger spring force to avoid the overlapping of two particles. Then the force on the $j$-th particle can be calculated by the Newton's third law.

## 3. Optimization of LBM calculation

The detail implementation of LBM is diverse though the evolution equation in Eq. (1) is simple and clear [19]. Here, we use the "push" scheme [12, 14] to implement the LBM calculation, in which the collision step and streaming step are done sequentially in each LB step. The standard algorithm to implement this calculation utilizes two arrays and two steps by divide Eq. (1) to two sequential steps as:

$$f_{p,\alpha}(\mathbf{x}, t + \Delta t) = f_\alpha(\mathbf{x}, t) - \frac{1}{\tau}\left[f_\alpha(\mathbf{x}, t) - f_\alpha^{(eq)}(\mathbf{x}, t)\right] + F_\alpha(\mathbf{x}, t)\Delta t \qquad (17)$$

and

$$f_\alpha(\mathbf{x} + \mathbf{e}_\alpha \Delta t, t + \Delta t) = f_{p,\alpha}(\mathbf{x}, t + \Delta t) \qquad (18)$$

in which Eq. (17) is the local collision step and Eq. (18) is the non-local streaming step. $f_{p,\alpha}$ is the additional temporary array for the post-collision PDFs, which are used for the non-local streaming operation.

The standard explicit time-marching operations within each time step exclude initialization and post-processing are as follows:

(1) Calculation of the local macroscopic flow quantities $p$ and $\mathbf{u}$ from the PDFs $f$.
(2) Calculation of the equilibrium distribution $f^{eq}$ from $p$ and $\mathbf{u}$ and execution of the "collision" (relaxation) operation to obtain post-collision PDFs $f^p$.
(3) Streaming, or someone called Propagation, of the $f^p$ to the neighboring nodes according to the discrete velocity direction $\alpha$ to obtain the new PDFs $f$ for the next time step calculation.

The first two steps are computationally intensive but involve only values of the local nodes while the third step is just a direction-dependent uniform shift of data in memory.

The collision-optimized array-of-structures (AoS) layout [19] is used for the data layout, in which the PDFs values of a node are assembled and stored consecutively in memory. Followed subsections are the detail algorithms for optimizations which are illustrated based on the D2Q9 model, they can be straightforwardly applied to other D$d$Q$q$ models.

### 3.1. Unrolling technique for LBM calculation

The standard calculation of fluid velocity in Eq. (5) consists a loop by nine multiplications. In fact, the discrete velocity includes many zero terms, which can be used for simplifying and reducing computation. Unroll the Eq. (5) by substituting the discrete velocity vectors in Eq. (2), we can get

$$u_x = (f_1 + f_2 - f_4 - f_5 - f_6 + f_8) \cdot c \tag{19}$$

$$u_y = (f_2 + f_3 + f_4 - f_6 - f_7 - f_8) \cdot c \tag{20}$$

In which only three additions, three subtractions and one multiplication are involved in each formula.

The collision calculation is the most time-consuming step for the whole LBM calculation. And then we can find that the calculations of the equilibrium distribution function (EDF) $f_\alpha^{(eq)}$ in Eq. (4) and the external term $F_\alpha(\mathbf{x}, t)$ in Eq. (8) are the main reasons. Similarly, the formula of equilibrium distribution function $f^{(eq)}$ in Eq. (4) and $F_\alpha(\mathbf{x}, t)$ in Eq. (8) can also be unrolled as

$$f_1^{(eq)} = 3\omega_1 \left[ p + \rho_0 \left( cu_x + u_x^2 - 0.5 u_y^2 \right) \right] / c^2 \tag{21}$$

$$f_2^{(eq)} = 3\omega_2 \left[ p + \rho_0 \left( cu_x + cu_y + 3 u_x u_y + u_x^2 + u_y^2 \right) \right] / c^2 \tag{22}$$

$$f_3^{(eq)} = 3\omega_3 \left[ p + \rho_0 \left( cu_y - 0.5 u_x^2 + u_y^2 \right) \right] / c^2 \tag{23}$$

$$f_4^{(eq)} = 3\omega_4 \left[ p + \rho_0 \left( -cu_x + cu_y - 3 u_x u_y + u_x^2 + u_y^2 \right) \right] / c^2 \tag{24}$$

$$f_5^{(eq)} = 3\omega_5 \left[ p + \rho_0 \left( -cu_x + u_x^2 - 0.5 u_y^2 \right) \right] / c^2 \tag{25}$$

$$f_6^{(eq)} = 3\omega_6 \left[ p + \rho_0 \left( -cu_x - cu_y + 3 u_x u_y + u_x^2 + u_y^2 \right) \right] / c^2 \tag{26}$$

$$f_7^{(eq)} = 3\omega_7 \left[ p + \rho_0 \left( -cu_y - 0.5 u_x^2 + u_y^2 \right) \right] / c^2 \tag{27}$$

$$f_8^{(eq)} = 3\omega_8 \left[ p + \rho_0 \left( cu_x - cu_y - 3 u_x u_y + u_x^2 + u_y^2 \right) \right] / c^2 \tag{28}$$

and

$$F_1(\mathbf{x},t) = 3\left(1 - \frac{1}{2\tau}\right)\omega_1 \left[ \frac{f_x}{c} + 3\frac{u_x f_x}{c^2} - \frac{u_x f_x + u_y f_y}{c^2} \right] \tag{29}$$

$$F_2(\mathbf{x},t) = 3\left(1 - \frac{1}{2\tau}\right)\omega_2 \left[ \frac{f_x + f_y}{c} + 3\frac{u_x f_y + u_y f_x}{c^2} + 2\frac{u_x f_x + u_y f_y}{c^2} \right] \tag{30}$$

$$F_3(\mathbf{x},t) = 3\left(1 - \frac{1}{2\tau}\right)\omega_3 \left[ \frac{f_y}{c} + 3\frac{u_y f_y}{c^2} - \frac{u_x f_x + u_y f_y}{c^2} \right] \tag{31}$$

$$F_4(\mathbf{x},t) = 3\left(1 - \frac{1}{2\tau}\right)\omega_4 \left[ \frac{-f_x + f_y}{c} - 3\frac{u_x f_y + u_y f_x}{c^2} + 2\frac{u_x f_x + u_y f_y}{c^2} \right] \tag{32}$$

$$F_5(\mathbf{x},t) = 3\left(1 - \frac{1}{2\tau}\right)\omega_5 \left[ \frac{-f_x}{c} + 3\frac{u_x f_x}{c^2} - \frac{u_x f_x + u_y f_y}{c^2} \right] \tag{33}$$

$$F_6(\mathbf{x},t) = 3\left(1 - \frac{1}{2\tau}\right)\omega_6 \left[ -\frac{f_x + f_y}{c} + 3\frac{u_x f_y + u_y f_x}{c^2} + 2\frac{u_x f_x + u_y f_y}{c^2} \right] \tag{34}$$

$$F_7(\mathbf{x},t) = 3\left(1-\frac{1}{2\tau}\right)\omega_7\left[-\frac{f_y}{c}+3\frac{u_y f_y}{c^2}-\frac{u_x f_x + u_y f_y}{c^2}\right] \tag{35}$$

$$F_8(\mathbf{x},t) = 3\left(1-\frac{1}{2\tau}\right)\omega_8\left[\frac{f_x - f_y}{c}-3\frac{u_x f_y + u_y f_x}{c^2}+2\frac{u_x f_x + u_y f_y}{c^2}\right] \tag{36}$$

where the $f_0^{(eq)}$ and $F_0$ is not need to be calculated because it's non-existed in the calculation of velocity and pressure.

*3.2. Symmetric technique for collision calculation*

Here, we use the symmetry trick to further optimize LB collision calculation. Firstly, we rearrange the expression of $f_\alpha^{(eq)}$ in Eq. (4) as

$$f_\alpha^{(eq)} = \omega_\alpha\left[\frac{p}{c_s^2} + \rho_0\left(\frac{(\mathbf{e}_\alpha \cdot \mathbf{u})^2}{2c_s^4} - \frac{|\mathbf{u}|^2}{2c_s^2}\right)\right] + \omega_\alpha \rho_0 \frac{(\mathbf{e}_\alpha \cdot \mathbf{u})}{c_s^2} \tag{37}$$

in which we can find that the first term on the right hand of the formula is equivalent for the two opposite discrete velocity directions $\alpha$ and $\alpha$-4 with $4<\alpha<9$. Hence $f_\alpha^{(eq)}$ ($4<\alpha<9$) can be calculated based on the $f_{\alpha-4}^{(eq)}$ as

$$f_\alpha^{(eq)} = f_{\alpha-4}^{(eq)} - 2\omega_{\alpha-4}\frac{(\mathbf{e}_{\alpha-4} \cdot \mathbf{u})}{c_s^2} \tag{38}$$

Fortunately, this trick can also be used for the calculation of the external force term in Eq. (8), which is the last term on the right hand of Eq. (1). The expression of $F_\alpha(\mathbf{x}, t)$ can also be rearranged as

$$F_\alpha(\mathbf{x},t) = \left(1-\frac{1}{2\tau}\right)\omega_\alpha\left[-\frac{\mathbf{u}(\mathbf{x},t)\cdot \mathbf{f}(\mathbf{x},t)}{c_s^2}+\frac{(\mathbf{e}_\alpha \cdot \mathbf{u}(\mathbf{x},t))(\mathbf{e}_\alpha \cdot \mathbf{f}(\mathbf{x},t))}{c_s^4}\right] + \left(1-\frac{1}{2\tau}\right)\omega_\alpha \frac{\mathbf{e}_\alpha \cdot \mathbf{f}(\mathbf{x},t)}{c_s^2} \tag{39}$$

in which on the right hand, the first term equals and the second term is opposite in the two opposite discrete velocity directions $\alpha$ and $\alpha$-4 with $4<\alpha<9$. Hence, $F_\alpha$ ($4<\alpha<9$) can be calculated based on the $F_{\alpha-4}$ as

$$F_\alpha(\mathbf{x},t) = F_{\alpha-4}(\mathbf{x},t) - 2\left(1-\frac{1}{2\tau}\right)\omega_{\alpha-4}\frac{\mathbf{e}_{\alpha-4} \cdot \mathbf{f}(\mathbf{x},t)}{c_s^2} \tag{40}$$

The second terms on the right hand of formula (38) and (40) for $f_\alpha^{(eq)}$ ($4<\alpha<9$) and $F_\alpha$ ($4<\alpha<9$) are calculated firstly in $f_\alpha^{(eq)}$ ($0<\alpha<5$) and $F_\alpha$ ($0<\alpha<5$). The Eq. (38) and (40) mean that only one subtraction between the calculation of $f_\alpha^{(eq)}$ ($0<\alpha<5$) and $F_\alpha$ ($0<\alpha<5$) with $f_\alpha^{(eq)}$ ($4<\alpha<9$) and $F_\alpha$ ($4<\alpha<9$). These indicate that this symmetry algorithm can approximately reduce half calculation amount of $f_\alpha^{(eq)}$ and $F_\alpha$.

*3.3. Swap algorithm for single f variable*

The standard implementation for Eq. (17) and (18) need to allocate an additional PDFs array $f^p[Nx+1][Ny+1][q]$ with equal memory size of basic PDFs array $f[Nx+1][Ny+1][q]$ for the streaming operation. Here, $Nx$ and $Ny$ are the number of lattices in the $x$ direction and $y$ direction. Therefore, the nodes are $Nx+1$ and $Ny+1$ in the $x$ direction and $y$ direction. The naive implementation can be seen in the Fig. 3(a) by staggered assign the value of array $fp$ to the array $f$. This operation is easy, while with double memory size of the PDFs, which cannot be suitable for a large-scale simulation. Even though the implementation of the two-lattice algorithm is straightforward, the extravagant usage of memory is unacceptable in the case of large lattice, and thus more economical implementations are needed.

Here, we use swap algorithm to eliminate the additional PDF array *fp* based on the work of [18] and [20]. In the swap algorithm the streaming step is implicitly performed by two swap operations, as can be seen in Fig. 3(b). The first swap is between the PDFs of the four directions ($\alpha$=1~4) and the PDFs of the opposite four directions ($\alpha$=5~8) on the local nodes, as can be shown in Fig. 1(a). And the second swap is between each PDF of the four directions ($\alpha$=1~4) on the target node and the according PDF of the opposite direction on the four adjacent nodes at the discrete lattice velocity direction. For example of the *y* direction streaming in Fig. 3(b), the first swap is swap(*f*[*x*][*y*][3],*f*[*x*][*y*][7]) on three nodes and the second swap is swap(*f*[*x*][*y*][3], f[*x*][*y*+1][7]) on the lower two nodes. Here, the swap operation exchanges their values. As can be seen that, the distribution of the PDFs in the Fig. 3. The standard streaming operation with double *f* variables (a) and the improved two-swap operation (local swap + adjacent swap) for single *f* variable (b). after two swap operations is the same with that in the Fig. 3(a) after one streaming operation. For programming code, Eq. (17) and (18) can be transformed as follows

$$f_\alpha(\mathbf{x},t) = (1-\omega)f_\alpha(\mathbf{x},t) \tag{41}$$

$$f_\alpha(\mathbf{x},t) = f_\alpha(\mathbf{x},t) + \omega f_\alpha^{(eq)}(\mathbf{x},t) + F_\alpha(\mathbf{x},t)\Delta t \tag{42}$$

In general, the code of a standard two-step and two arrays algorithm can be simply modified by substituting the streaming step using the two-swap steps. For the cache-based optimization, the first swap can be coupled to the collision calculation process. We can see that the memory of PDFs is reduced from 2×(*Nx*+1)×(*Ny*+1)×*q* to (*Nx*+1)×(*Ny*+1)×*q*, while the complexity of this algorithm does not increase. Another advantage is that this process is conservative and no information is ever destroyed. It has no problems for the parallel computing. In addition, there is a bit point should be state: the first local swap is symmetry while the second non-local swap is non-symmetry but directional. Mattila et al. [18] indicate that the PDFs of the boundary must be treated separately after the whole lattice has been updated. Latt [20] swap PDFs on all nodes including the boundary node while restrict all the adjacent nodes inside the computational domain. In fact, the number of the second swaps can be *Ny* at the y direction, which is one less than the nodes number, as can be seen in Fig. 3(b). Due to the directionality of the second swap, we can change coordinate *y* from 0 to *Ny*-1 by swap(*f*[*x*][*y*][$\alpha$], f[*x*+e[$\alpha$][0]][*y*+e[$\alpha$][1]][$\alpha$+4]) with 1 ≤ $\alpha$ ≤ 4 or from 1 to *Ny* by swap(*f*[*x*][*y*][$\alpha$-4], *f*[*x*+e[$\alpha$][0]][*y*+e[$\alpha$][1]][$\alpha$]) with 5≤$\alpha$≤8. Here, we choose the former and so does the coordinate *x*. The PDFs on the boundary nodes are not need to be treated particularly.

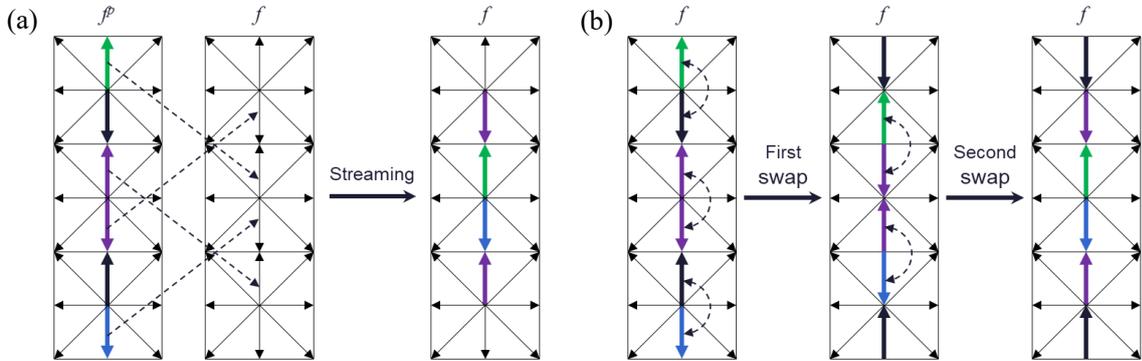

Fig. 3. The standard streaming operation with double *f* variables (a) and the improved two-swap operation (local swap + adjacent swap) for single *f* variable (b).

In summary, above improved algorithms for the implementation of LBM calculation based on C programming language are depicted in Table 1. It can be seen that the code is concise, which will be expected much more efficient than the standard code.

**Table 1: Improved LBM algorithm**

```
for (x=1; x< Nx; x++)
for (y=1; y< Ny; y++)
  p[x][y] =…;
  u[x][y][0] =…;                                        //Calculate the macro velocity and pressure by Eq. (6), (19)
  u[x][y][1] =…;                                        and (20)

for (x=0; x<=Nx; x++)
for (y=0; y<=Ny; y++)
  f^eq[i] =…;       (0<i<9)                             //Calculate the EDF by Eq. (21)-(24) and Eq. (38)
  F_IB[i] =…;       (0<i<9)                             //Calculate external force term by Eq. (29)-(32) and Eq. (40)
  f[x][y][i] *= (1 - ω);
  f[x][y][i] += w[i] * (ω* f^eq[i] + (1-0.5 *ω) * F_IB[i] * dt);    //Collision calculation by Eq. (41) and (42)
  for (i=1; i<=4; i++)
                                                        //First swap on the local node
    swap(f[x][y][i], f[x][y][i+4]);
for (x=0; x<Nx; x++)
for (y=0; y<Ny; y++)
  for (i=1; i<=4; i++)                                  //Second swap between adjacent nodes in the interior area
    xd = x + e[i][0];                                   excluding the boundary
    yd = y + e[i][1];
    swap(f[x][y][i], f[xd][yd][i+4]);
```

## 4. Local point-to-node algorithm for IBM optimization

There is fewer literatures for the detail implementation of the IBM calculation for particle-laden flows. For solving the interaction between particles and fluids, the detail computing procedure consists three sequential steps: (1) search effective fluid neighbor nodes of each Lagrangian points and interpolate of fluid velocity from fluid nodes to the Lagrangian points, (2) calculate the immersed boundary (IB) force by the difference between Lagrangian velocity and the interpolation fluid velocity, (3) search effective neighbor Lagrangian points of each fluid nodes and spread IB force from Lagrangian points to the fluid nodes. The first step and the third step are the most-time consuming. The naive implementation of these two steps are

$$\mathbf{u}_b^*(\mathbf{X}_k) = \sum_{i=0}^{Nx}\sum_{j=0}^{Ny} \mathbf{u}^*(\mathbf{x}_{i,j})\delta(\mathbf{x}_{i,j} - \mathbf{X}_k), \quad 0 \leq k < NL \tag{43}$$

$$\mathbf{f}(\mathbf{x}_{i,j}) = \sum_{k=0}^{NL} \mathbf{F}(\mathbf{X}_k)\delta(\mathbf{x}_{i,j} - \mathbf{X}_k)\Delta s, \quad 0 \leq i \leq Nx, 0 \leq j \leq Ny \tag{44}$$

Here, the whole fluid nodes and the whole boundary points are searched. Its operation count is $N_0 = O(2N_L \times (Nx+1) \times (Ny+1))$, which is too expensive to be used for numerical simulation. As the depiction in Section 2.2, the Lagrangian points on the particle boundary only interact with the neighborhood fluid nodes due the using of the regularized delta function. In fact, too many fluid nodes and boundary points are searched in the naive implementation, while not in the effective range. The standard algorithm reduces the search area of fluid nodes to a small box surrounding the particle, as can be seen in Fig. 4(a). The small box should be slightly larger than the particle with area $(Nx2-Nx1) \times (Ny2-Ny1)$, in which $Nx1=\text{int}(xc-R-r)$, $Nx2=\text{int}(xc+R+r)$, $Ny1=\text{int}(yc-R-r)$ and $Ny2=\text{int}(yc+R+r)$. Then the Eq. (43) and (44) can be updated as

$$\mathbf{u}_b^*(\mathbf{X}_k) = \sum_{i=Nx1}^{Nx2} \sum_{j=Ny1}^{Ny2} \mathbf{u}^*(\mathbf{x}_{i,j}) \delta(\mathbf{x}_{i,j} - \mathbf{X}_k), \quad 0 \leq k < NL \tag{45}$$

$$\mathbf{f}(\mathbf{x}_{i,j}) = \sum_{k=0}^{NL} \mathbf{F}(\mathbf{X}_k) \delta(\mathbf{x}_{i,j} - \mathbf{X}_k) \Delta s, \quad Nx1 \leq i \leq Nx2, Ny1 \leq j \leq Ny2 \tag{46}$$

The programming implementation of this standard algorithm can be seen in Table 2. Two large loops are used for the velocity interpolation operation of Eq. (45) and the force spreading operation of Eq. (46), respectively. It should be point out that the IB force calculation is fused in the former loop. Its operation count is reduced to $N_1=O(2N_L \times (Nx2-Nx1) \times (Ny2-Ny1))$, which is clearly much less than $N_0$. However, we find that there are still many fluid nodes are searched, while not in the effective range, such as particle interior part and the four corner parts in the Fig. 4(a). And for each fluid nodes, all the Lagrangian points should be searched, while most of them are out the effective range. Now, we propose to narrow the box to be the smallest surrounding each Lagrangian boundary points to eliminate any unnecessary search operation, as can be seen in Fig. 4(b). The smallest box size can be only the effective range of a single delta function with $(Nx4-Nx3) \times (Ny4-Ny3)$, in which $Nx3=\text{int}(X[k]-r)$, $Nx4=\text{int}(X[k]+r)$, $Ny3=\text{int}(Y[k]-r)$ and $Ny4=\text{int}(Y[k]+r)$. $r$ is the effective range of the delta function and equal to be $1.5dh$ for the 3-point delta function used in this paper. Hence, both the velocity interpolation in the first step and the IB force spreading in the third step are calculated by point-to-node. Then the Eq. (45) can be further updated as

$$\mathbf{u}_b^*(\mathbf{X}_k) = \sum_{i=Nx3}^{Nx4} \sum_{j=Ny3}^{Ny4} \mathbf{u}^*(\mathbf{x}_{i,j}) \delta(\mathbf{x}_{i,j} - \mathbf{X}_k), \quad 0 \leq k < NL \tag{47}$$

Then the force on the fluid nodes in the local smallest box can be obtained by

$$\mathbf{f}(\mathbf{x}_{i,j}) = \mathbf{F}(\mathbf{X}_k) \delta(\mathbf{x}_{i,j} - \mathbf{X}_k) \Delta s, \quad Nx3 \leq i \leq Nx4, Ny3 \leq j \leq Ny4 \tag{48}$$

It should be noted that the final force exerted on a fluid node should be superposed by above term of neighbor Lagrangian points in the effective range. It is clearly that, this operation count in this point-to-node algorithm is reduced to $N_2=O(2N_L \times (Nx4-Nx3) \times (Ny4-Ny3)) = O(8N_L r^2)$ in two dimensional simulations.

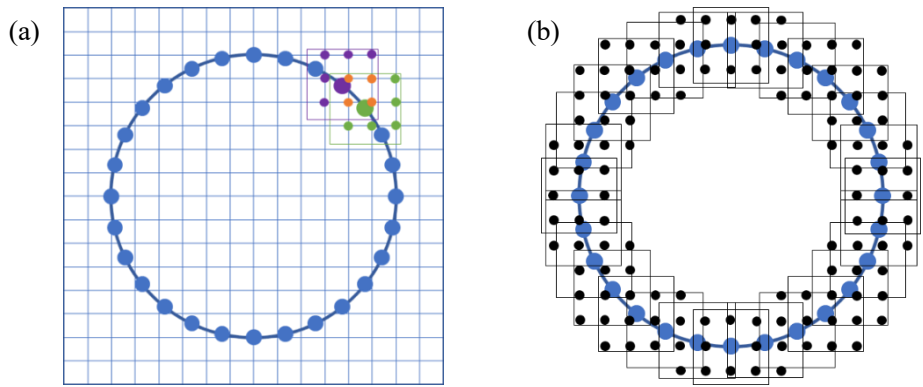

Fig. 4. The interaction between solid Lagrangian points and fluid Eulerian nodes. (a) standard algorithm and (b) improved point-to-node algorithm.

Firstly, we quantitatively compare the difference between $N_1$ and $N_2$ depending on the particle diameter resolution $D_p/dh$. Here, the 3-point delta function is used and $r=1.5dh$. As can be seen in Fig. 5(a) and (c), $N_1$ increases much more sharply than $N_2$ when the particle diameter resolution increases. When $D_p/dh=10$ for the low Reynolds number flows, $N_1$ are 20 and 80 times of $N_2$ in two- and three-dimensional cases. While when $D_p/dh=40$ for the high Reynolds number flows, $N_1$ become 200 and 3000 times of $N_2$ in two- and three-dimensional cases, respectively. In a logarithm coordinate in Fig. 5(b) and (d), we can find that all these curves are in exponent. The index number of $N_1$ is much larger than that of $N_2$, whether in two- or three-dimensional cases.

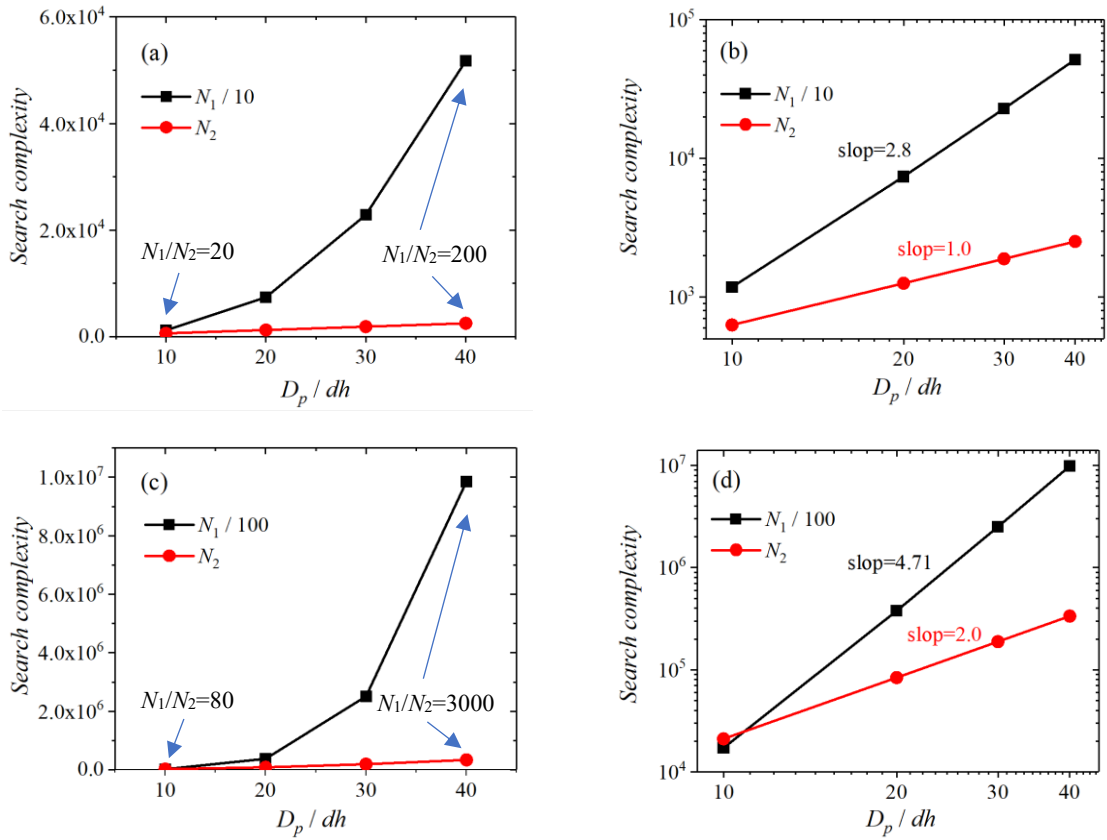

Fig. 5. The search complexity of IBM calculation. (a) and (b) are the two-dimensional case while (c) and (d) are the three-dimensional case, (a) and (c) are in linear coordinate while (b) and (d) are logarithm coordinate. Black square and red circle depict the standard algorithm and the proposed point-to-node algorithm, respectively.

**Table 2: Standard IBM algorithm for particle-laden flows**

| Code | Comment |
|---|---|
| for (i=0; i< $N_L$; i++) | //loop through all the Lagrangian points |
|   xb[i]=xc+Rcos$\theta$;  yb[i]=xc+Rcos$\theta$; | //Initialization of the coordinate, solid velocity and the unforced fluid velocity at the Lagrangian points on the particle boundary. |
|   *up*[i][0]=…;  *up*[i][1]=…; | |
|   *uf*[i][0]=0.0;  *uf*[i][1]=0.0; | |
|   for (x=int (xc-R-r); x<= int (xc+R+r); x++) | //loop through the local fluid nodes around the particle boundary |
|   for (y=int (yc-R-r); y<= int (yc+R+r); y++) | |
|     lx=x-xb[i];  ly=y-yb[i]; | //Calculate the unforced fluid velocity at the Lagrangian points by interpolation |
|     *uf*[i][0] += *u*[x][y][0]*delta(lx)* delta(ly); | |
|     *uf*[i][1] += *u*[x][y][1]*delta(lx)* delta(ly); | |
|   F_IB[i][0]=2*$\rho$*( *up*[i][0]- *uf*[i][0])/dt; | //Calculate the immersed boundary force at the Lagrangian points |
|   F_IB[i][1]=2*$\rho$*( *up*[i][1]- *uf*[i][1])/dt; | |
| | |
| for (x=int (xc-R-r); x<= int (xc+R+r); x++) | //loop through the local fluid nodes around the particle boundary |
| for (y=int (yc-R-r); y<= int (yc+R+r); y++) | |
|   for (i=0; i< $N_L$; i++) | //loop through all the Lagrangian points |
|     lx=x-xb[i];  ly=y-yb[i]; | //Calculate the immersed boundary force on the fluid nodes by spreading |
|     f_IB[x][y][0]+= F_IB[i][0] *delta(lx)* delta(ly)*ds*drs; | |
|     f_IB[x][y][1]+= F_IB[i][1] *delta(lx)* delta(ly)*ds*drs; | |

**Table 3: Improved point-to-node IBM algorithm**

| Code | Comment |
|---|---|
| for (i=0; i< $N_L$; i++) | //loop through all the Lagrangian points |
|   xb[i]=xc+Rcos$\theta$;  yb[i]=xc+Rcos$\theta$; | //Initialization of the coordinate, solid velocity and the unforced fluid velocity at the Lagrangian points on the particle boundary. |
|   *up*[i][0]=…;  *up*[i][1]=…; | |
|   *uf*[i][0]=0.0;  *uf*[i][1]=0.0; | |
|   for (x=int (xb[i]-r); x<= int (xb[i]+r); x++) | //loop through the local fluid nodes adjacent the target point |
|   for (y=int (yb[i]-r); y<= int (yb[i]+r); y++) | |
|     lx=x-xb[i];  ly=y-yb[i]; | //Calculate the unforced fluid velocity at the Lagrangian points by interpolation |
|     *uf*[i][0] += *u*[x][y][0]*delta(lx)* delta(ly); | |
|     *uf*[i][1] += *u*[x][y][1]*delta(lx)* delta(ly); | |
|   F_IB[0]=2*$\rho$*( *up*[i][0]- *uf*[i][0])/dt; | //Calculate the immersed boundary force at the Lagrangian points |
|   F_IB[1]=2*$\rho$*( *up*[i][1]- *uf*[i][1])/dt; | |
| | |
|   for (x=int (xb[i]-r); x<= int (xb[i]+r); x++) | //loop through the local fluid nodes adjacent the target point |
|   for (y=int (yb[i]-r); y<= int (yb[i]+r); y++) | |
|     lx=x-xb[i];  ly=y-yb[i]; | //Calculate the immersed boundary force on the fluid nodes by spreading |
|     f_IB[x][y][0]+= F_IB[0] *delta(lx)* delta(ly)*ds*drs; | |
|     f_IB[x][y][1]+= F_IB[1] *delta(lx)* delta(ly)*ds*drs; | |

## 5. Grid search algorithm for particle-pair force calculation

For the particle-pair force calculation in a large-scale system, the most expensive part is the neighbor search. First, we introduce the half search algorithm here, which also has been improved from the naive direct search algorithm, or called "all-in-all" algorithm. In the "all-in-all" algorithm, all other particles should be searched for each particle. Then, the operation count is $N_p(N_p-1)$, which is the most time-consuming. And it has another problem that each contact is double computed, which is further more time-consuming. In fact, each contact force can be computed only once and the force applied on another particle can be the directly obtained by the Newton's third law. The half search algorithm only searches the rest particles whose index larger than the target particle while not all other particles. Then operation count can be reduced to $0.5N_p(N_p-1)$, while is also $O(N_p^2)$.

**Table 4: Half-search algorithm for particle-pair interaction**:

| | |
|---|---|
| **for** ($i = 0$; $i < N_p$; $i$++) | //For each object particle |
|   **for** ($j = i+1$; $j < N_p$; $j$++) | //Search the particles with label bigger than it |
|     if ($L_{jk} - D_p < \zeta$)  $F_{lub}$=… | //Calculate lubrication force on particle $i$ by Eq.(16). |
|       $F_{lub}[i]$ += $F_{lub}$ | |
|       $F_{lub}[j]$ += $-F_{lub}$ | |

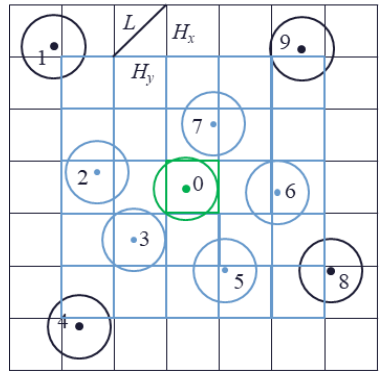

Fig. 6 Illustration of grid search algorithm for particle-pair short-range interaction.

As we all know, a particle can only contact with the few neighbor particles (no more than six particles in two-dimensional case) at a certain time. Based on this, a well-known efficient algorithm is called linked-cell method which is originated from Molecular Dynamics simulation [36] and then applied to Computational Granular Dynamics [33] and CFD-DEM simulation [47]. The method divides the computational domain into small rectangular boxes, so-called "cells", which is slightly larger than particle diameter at any dimension. This method can reduce the complexity to $O(N_p)$ because the search region for each particle has been narrowed to the neighbor 8 two-dimensional cells or 26 three-dimensional cells of the target particle. However, this method still exists a bit problem that there may be several particles in each cell. The particles shared in a common cell should be distinguished, which conversely increases the calculation complexity. Here, we adopted the grid search algorithm [39] which is proposed based on the linked-cell algorithm. The difference is that the cell size is determined by the requirement that each cell hosts no more than one particle. It is high efficient, more easier to implementation [33]. Here, this method is further clearer that the size of the cell is determined by two conditions. The first condition is that each cell can contain the center of no more than one particle. Hence the diagonal length $L$ of a cell should be shorter than the particle diameter as

$$L = \sqrt{H_x^2 + H_y^2} < D_p \tag{49}$$

in which $H_x$ and $H_y$ is the cell step in x and y direction, respectively. That means particles and cells will be mapped each other point to point. Since continuous and random positions of the particles are mapped into the cells, the specification of neighbor particles becomes very easy and pointers for neighbor sub-cells or link list are not needed yet. However, the cell size cannot be infinitely small because of computational efficiency. The second condition is that only the most nearby 25 cells (2D) and 125 cells (3D) (including the one hosting the particle) are detected during particle-pair force calculation, as shown in Fig. 6. Hence the edge length of the cells should be larger than half of the effective force range as

$$2H_x(H_y) > D_p + \zeta \tag{50}$$

in which $\zeta$ is the threshold or cutoff distance of the particle-pair lubrication force in Eq. (16). Eq. (49) and (50) decide the maximum size and the minimum size of the cell, respectively. If $H_x = H_y = H$ (i.e. the cell is square), we can get that $0.5D_p < H < 0.707D_p$. And in the same way we can get $0.5D_p < H < 0.577D_p$ for the three-dimensional calculations.

By this way, all particles can be completely mapped into each cell, i.e., one cell corresponds to one particle or not, respectively.

$$I[x][y] = \begin{cases} k & \text{if cell contains the center of particle } k \\ -1 & \text{if cell contains no particle} \end{cases} \tag{51}$$

It should be pointed out that there exist two real search marching strategies: search marching as particle index or search marching as grid cell index because they are corresponded to particles point-to-cell. We select the former one as clearly the number of particles is less than the number of the cells. The simulation algorithm is then described by the steps in Table 5.

**Table 5: Grid search algorithm for particle-pair interactions**:

| | |
|---|---|
| $Value[X][Y] = -1$; | Initial evaluation for all particle cells, value -1 means no occupation of particles |
| **while** $k \leq N_p$ **do**<br>    $X_p = \text{int}(x_p / H_x)$;    $Y_p = \text{int}(x_p / H_y)$;<br>    $Value[X_p][Y_p] = k$;<br>**end** | The mutual mapping between cells and particles, the particle $k$ location of the cell is $X_p$ and $Y_p$, while the Value of this cell maps the particle index $k$. |
| **while** $k \leq N_p$ **do**<br>    **for** $(X = X_p - 2; X \leq X_p + 2; X++)$<br>    **for** $(Y = Y_p - 2; Y \leq Y_p + 2; Y++)$ | Search the neighboring 25 cells |
|       **if** $(Value[X][Y] \neq -1$ && $Value[X][Y] \neq k)$ | The potential interaction particle judge determined by the cell Value and remove itself. |
|         $j = Value[X][Y]$;<br>        **if** $(L_{jk} - D_p < \zeta)$    $F_{lub}[k] = -F_{lub}[j] = $ Eq.(16); | The calculation of lubrication force between particle $k$ and particle $j$. |
|     $Value[X_p][Y_p] = -1$;<br>**end** | Remove this particle from the cell for the following calculations. |

## 6. Performance results and discussions

In order to test the performances, a well-known problem in particle-laden flows, particles settling in a closed cavity [31, 45] is used to compare the present improved algorithms with the standard algorithms. Here, the improved algorithms and the standard algorithms are defined in Table 6, respectively. Both the two-dimensional (2D) and three-dimensional (3D) flows and both the particle-free and particle-laden flows are involved. Simulations are performed on a computer cluster equipped with Intel Xeon E6240 (2.60 GHz, 18 cores) CPU processors. Each node has 2 CPU processors and 6 chips of 16GB RAM. The operation system is GNU/Linux X86_64, SMP 2.6.18-194.e15 and our program codes are written in ANSI C and compiled by gcc complier with optimization command -Ofast. The simulations are run for 1000 time steps in two-dimensional case and 200 time steps in three-dimensional case for statistics. The aim of our simulations is to compare the whole performances and the parts performances of algorithms under the effects of the particle volume fraction and the particle diameter resolution. We firstly consider the performance on the particle-free flows and then the particle-laden flows.

Table 6: The definition of the standard algorithm and the improved algorithm for comparing.

| Item | LBM | IBM | Particle-pair collision |
| --- | --- | --- | --- |
| Standard algorithm | Standard | Standard | Half-search |
| Improved algorithm | Unrolling, Symmetric and Swap | Point-to-node | Grid-search |

### 6.1 Particle-free flows

For the particle-free flows, the closed 2D and 3D cavities are transferred to two Poiseuille flows by changing the bottom boundary and top boundary conditions to be inflow and outflow conditions, respectively. The boundary conditions of the fluid domain are implemented by the non-equilibrium extrapolation method [48]. Several simulations are performed with varied square and cube domain size from $40^2$ up to $4000^2$ for 2D simulation and $40^3$ up to $210^3$ for 3D simulation. The performances of the algorithms are depicted by MLUPS (Mega Lattice Site Update per Second), which is handy unit for measuring the performance of LBM. It allows an easy estimation of the running of a real application depending on the domain size and the number of desired time steps as

$$MLUPS = \frac{N_{Grid} \times N_t}{10^6 \times t} \qquad (52)$$

where $N_{Grid}$ and $N_t$ are the number of fluid grids and the number of computational time steps, and $t$ is the total real time for the simulations.

Fig. 7 demonstrates the performances of the standard and improved implementations of the incompressible LBM for particle-free flows. We can see that the MLUPS values tend to be constant after early increasing stage, which is affected by the boundary implementation of the non-equilibrium extrapolation method. The MLUPS values obtained by the improved algorithm are much higher than those obtained by the standard algorithm, especially in 3D cases. For example, the approximately MLUPS values in Fig. 7(a) are ~21 and ~36 for the standard algorithm and the improved algorithm when $N_{grid} \geq 1000^2$. This means that the present algorithm improves the performance of the program by about 1.7 times compared to the standard algorithm.

While the MLUPS values in the 3D cases are generally lower than in the 2D cases, the improved effects are qualitatively much better, as shown in Fig. 7 (b). The approximately MLUPS values are 5.2 and 12 for the standard algorithm and the improved algorithm when $N_{grid} \geq 80^3$. Hence, the speedup ratio of the latter to the former is about 2.3. In order to compare with previous literatures, we find that the MLUPS values in literature [49] are 7.42 and 9.57 for domain sizes of $128^3$ and $192^3$. It should be noted that these improvements are not based on cache optimization and hence the performances are generally in any computer machine.

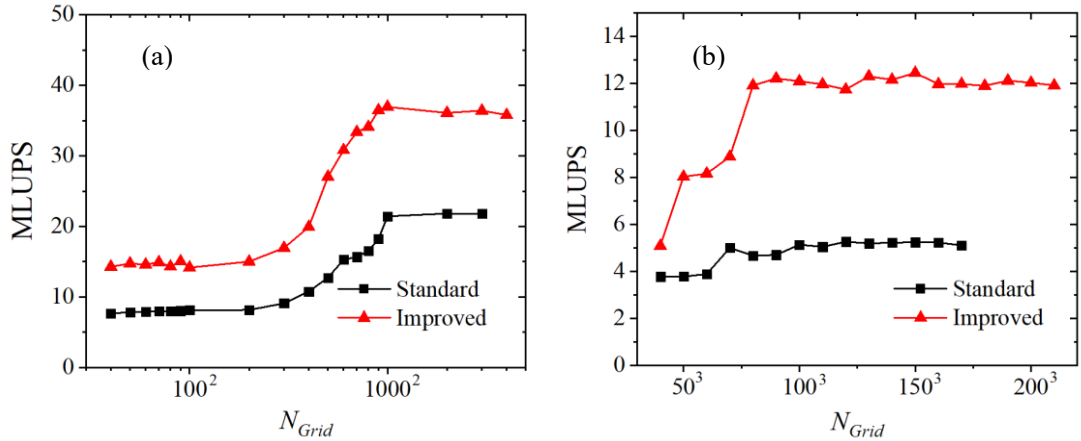

Fig. 7. Performance with lattice site update rate of the algorithms. (a) two dimensional and (b) three dimensional.

6.2 Particle-laden flows

For the particle-laden flows, slightly heavier particles ($\rho_p / \rho_f = 1.01$) are settled in initially rest fluid in the closed cavities from initially random positions, as can be seen in Fig. 8. The ambient boundaries of the fluid domain are fixed and also implemented by the non-equilibrium extrapolation method. 3-point delta function and the corresponding parameter $drs/dh=1.9$ for IB calculations are used in this part. In order to measure the performance on LBM, IBM and particle-pair collisions, we have executed two series of simulations. First is the effect of particle volume fraction and particle number in fixed fluid domain. And second is the effect of particle resolution ($D_p/dh$) in fluid domain with fixed volume fraction are considered respectively.

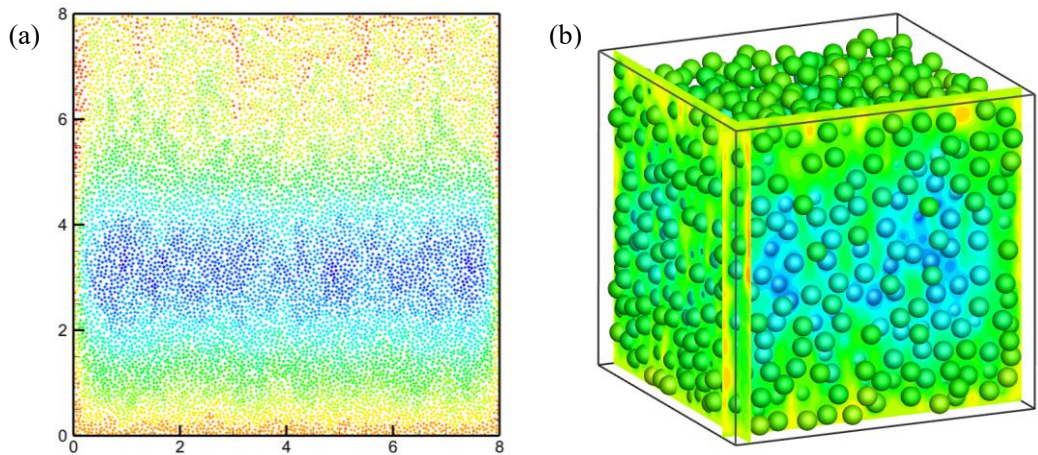

Fig. 8 Simulation of (a) 9400 2D circle particles and (b) 2000 3D sphere particles settling in closed cavities with initial random distribution by a single CPU core. The particle color depicts the $y$-velocity.

6.2.1 Effect of particle volume fraction

In this section, the 2D and 3D fluid domains are fixed as $2048^2$ 2D grids and $150^3$ 3D grids, and the number of particles is varied from 1 to $10^4$ according to the former, and from 1 to 600 according to the latter. The simulations are run for 1000 time steps for the 2D case, and 200 time steps for the 3D case. The detail parameters can be shown in Table 7. The mesh ratio $\eta$ is the ratio of the number of total Lagrangian points to the total fluid Eulerian nodes as $\eta=N_p*N_L/N_{Grid}$, in which $N_L$ is the number of Lagrangian points on a single particle's surface. LB calculation dependences on the number of fluid nodes while IB calculation dependences on the number of Lagrangian boundary points.

Table 7. Calculation parameters for the 2D and 3D tests.

| Dimensions | Fluid grids $N_{Grid}$ | Number of particles $N_p$ | Particle volume fraction $V_p$ | Particle resolution $D_p/dh$ | Mesh ratio $\eta$ |
|---|---|---|---|---|---|
| 2D | $2048^2$ | 1~$10^4$ | 0.0048%~48% | 16 | 0.0013%~13.4% |
| 3D | $150^3$ | 1~600 | 0.027%~27% | 12 | 0.01%~6.9% |

Fig. 9 illustrates the contour of IB force distributed on the Eulerian grids in the fluid field. The 2D circle particles in Fig. 9(a) and 3D spheres in Fig. 9(b) are discreted by Lagrangian points uniformly distributed on their surfaces. In the supplementary video S1 and S2, the Lagrangian points move following particles moving. It can be clearly seen that the IB forces only exist in the nearby area around the Lagrangian points on the particle boundaries.

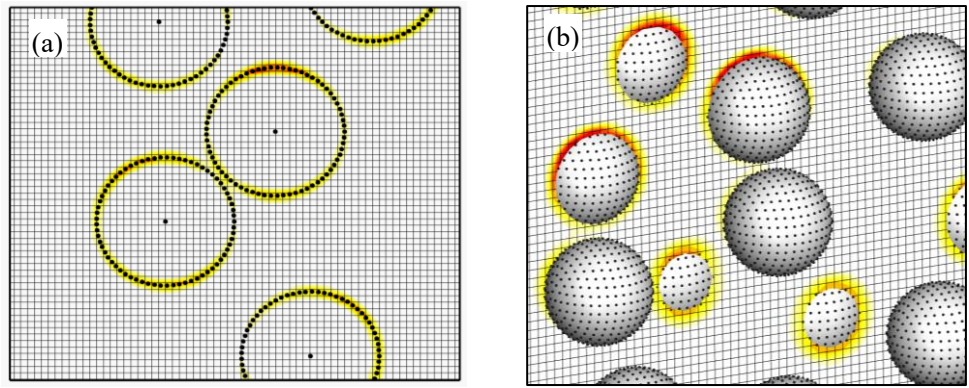

Fig. 9. Illustration of local IB force on the Eulerian grids around the Lagrangian points on the particle boundaries. (a) 2D and (b) 3D.

We calculate the time consumption of each part of the simulation: macro quantity calculation by LB, Boundary implementation of fluid domain, IB calculation, LB collision, LB streaming, particle collision and particle movement. Fig. 10 shows the time consumption of each part except the particle movement because the time consumption of this part is much lower for only 1000 particles in 2D and 100 particles in 3D. We can find almost all parts are dramatically decreased for the improved algorithm compared with the standard algorithm, especially the part of IBM calculation. The total calculation time of the improved algorithm to that of the standard algorithm is 60% in 2D and 25% of in 3D, respectively.

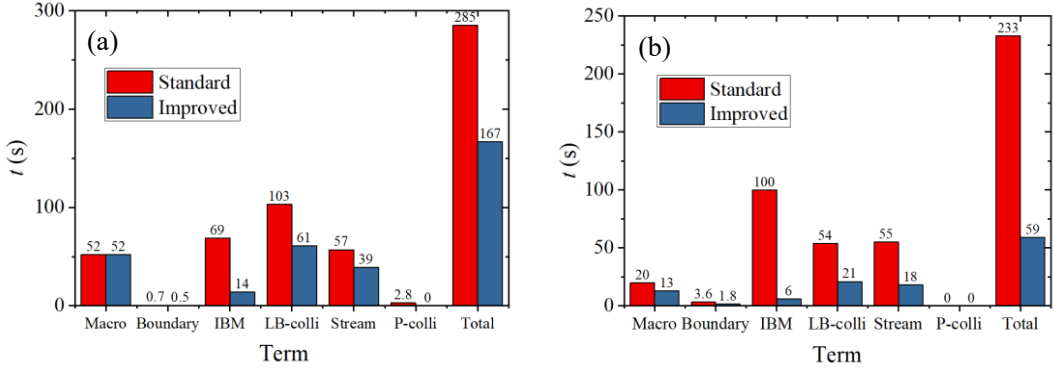

Fig. 10. Calculation time of each part. (a) 2D test with $N_{Grid}=2048^2$, $N_p=1000$, $D_p/dh=16$ and $N_t=1000$; (b) 3D test with $N_{Grid}=150^3$, $N_p=100$, $D_p/dh=12$ and $N_t=200$.

Fig. 11 shows the time consumption of IB calculation varies with the particle's number and volume fraction in a fixed fluid domain for the particle-laden flows. Firstly, we can find almost all lines in Fig. 11(a) and (b) are straight with slope equaling to about one. That means the amount of IBM calculation linearly varies with the particle number and the particle volume fraction. Secondly, the time consumption of the improved algorithm is much less than that of the standard algorithm. The difference is due to the omitted of the unnecessary node searching by local point-to-node algorithm. This decreasing effect of the 3D test is more obvious because more nodes searching omitted is involved.

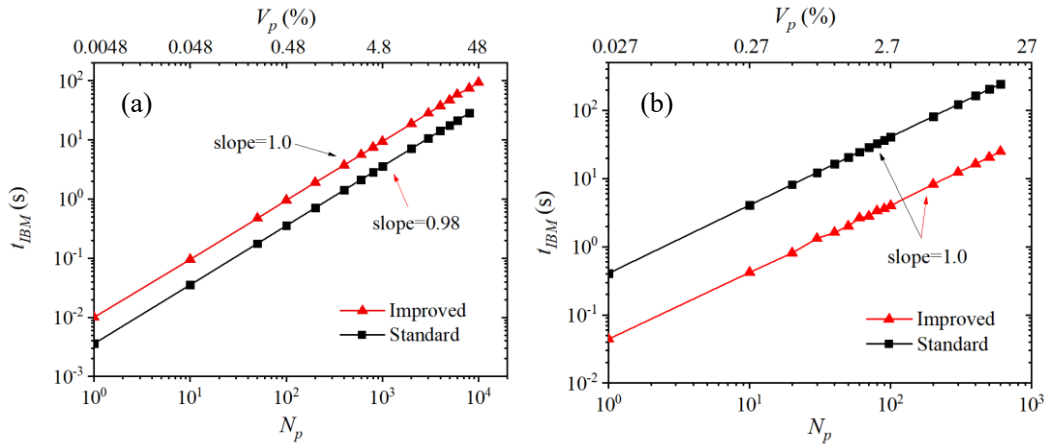

Fig. 11. Calculation time of IB for particle-laden flows. (a) 2D test (fluid domain $2048^2$, particle resolution is 16), (b) 3D test (fluid domain $150^3$, particle resolution is 12).

The coupled IB-LBM method here affects the fluid field by a spitting IB force scheme: one half IB force is used to directly correct the fluid velocity and another half IB force is used to correct the PDFs by the external force term in the collision operation in Eq. (1). Fig. 12 shows the time consumption of LB collision varies with particle number in a fixed fluid domain. It can be seen that they all increase as the particle number $N_p$ increases. However, the increase rate is low but slightly higher for the larger particle volume fraction i.e. larger number of particles. The time consumption of the LB collision obtained by the improved algorithm is much less than that obtained by the standard algorithm. This is mainly due to the unrolling and symmetric algorithms for the calculation of the equilibrium distributed functions in the external force term.

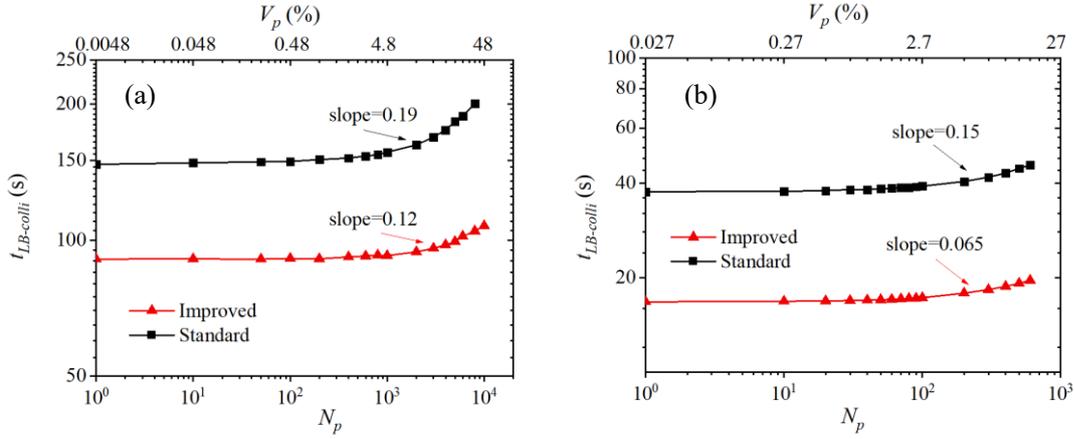

Fig. 12. Calculation time of LB collision for particle-laden flows. (a) 2D test, (b) 3D test (fluid domain $150^3$, particle resolution is 12).

The number of particles is up to $10^4$ and 600 in 2D and 3D, respectively, due to the computer capacity limitation. And the time consumption of particle-pair collision based on particle number is shown in Fig. 13. We can see that when particle number is small, for example $N_p$<100 in 2D and $N_p$<50 in 3D, the time consumption obtained by the improved algorithm is higher than that obtained by the standard algorithm. This is due to the time consuming of the initialization of the search grids and the mutual mapping between particles and the search grids before the search of the particle-pair collisions. When particle number is large, it is obvious that the standard algorithm causes the calculation time increasing as $O(N_p^2)$ while the improved algorithm brings linearly increase.

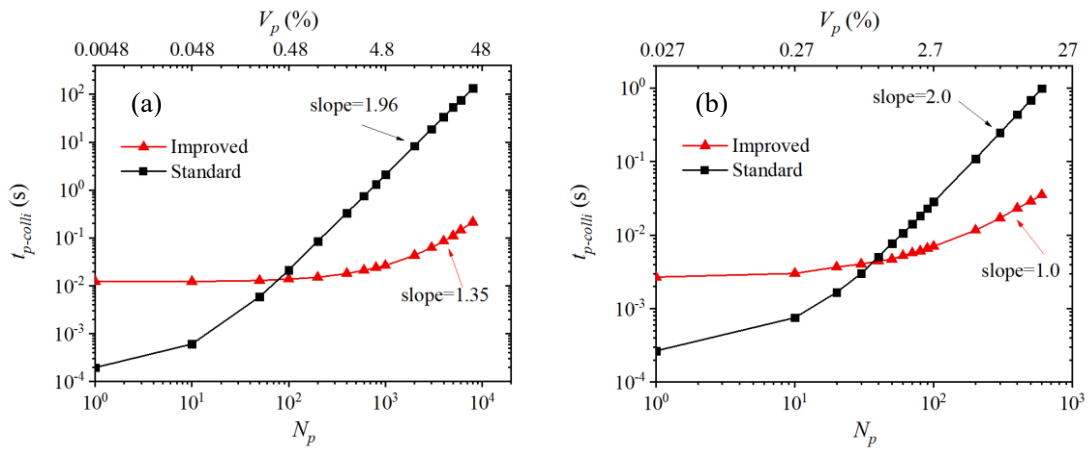

Fig. 13. Calculation time of particle-pair collision for particle-laden flows. (a) 2D test, (b) 3D test (fluid domain $150^3$, particle resolution is 12).

Fig. 14 shows the total time consumption of the whole computing. For the dilute flow with low particle volume fraction, the curves are more similar to those of LB collision computing in Fig. 12. However, for the dense flow with high particle volume fraction, the curves become similar to those of IB computing in Fig. 11. The slops of all the curves are lower than one and those obtained by the improved algorithm are about half of those obtained by the standard algorithm.

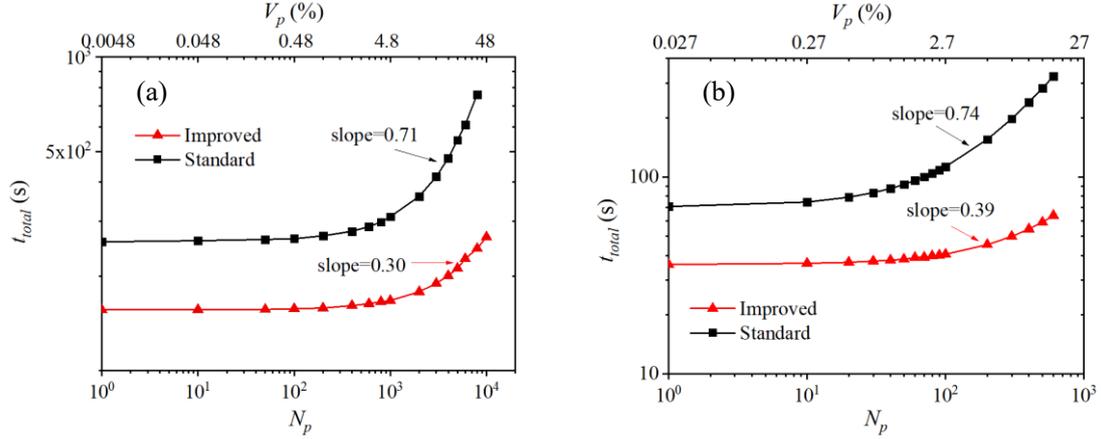

Fig. 14. Calculation time of whole computing for particle-laden flows. (a) 2D test, (b) 3D test (fluid domain $150^3$, particle resolution is 12).

We define a relative efficiency parameter for the whole calculation as $\beta = t_{standard} / t_{improved}$, which is drawn in Fig. 15. It can be seen that $\beta$ is larger than 1.5 and 2.0 for 2D and 3D fluid flow even with only one particle. $\beta$ increases slightly when particle number is small while dramatically when particle number becomes large. For example, $\beta$ equals to 4.8 when particle number is 500 in 3D test. This indicates that the improved algorithm has much higher efficiency compared with the standard algorithm. Higher volume fraction with larger number particles in a fixed fluid domain can reflect higher efficiency.

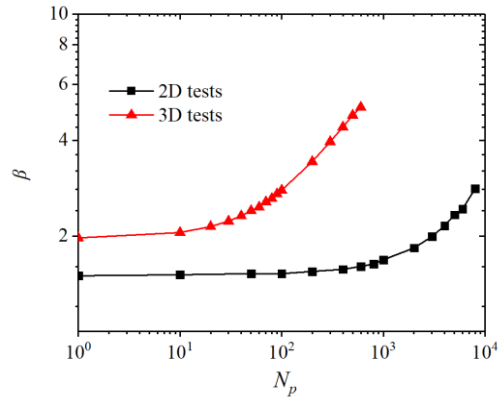

Fig. 15. The performance efficiency of the improved algorithm compared to the standard algorithm for particle-laden flows.

Fig. 16 illustrates the percent of each calculation parts vary with particle number $N_p$ and the mesh ratio $\eta$. Here, the mainly part including LB macro quantity calculation, IBM calculation, LB-collision, particle-pair collision are considered. Others includes the LB streaming and particle movement, etc. It can be seen that the calculation of LB collision dominates the whole calculation when the particle number is small. For the standard algorithm, the IBM calculation gradually dominates when particle number increases while the percent of LB collision is gradually decreased. For example, the percent value of IBM calculation increases to 41% and 76% while that of LB collision decreases to 26% and 14% for the particle number of $10^4$ and 600 in 2D and 3D flows, respectively. But for the improved algorithm, the time distribution of each part is more even. The LB-collision part and the IBM part are the two mainly part for the whole calculation when particle number is large.

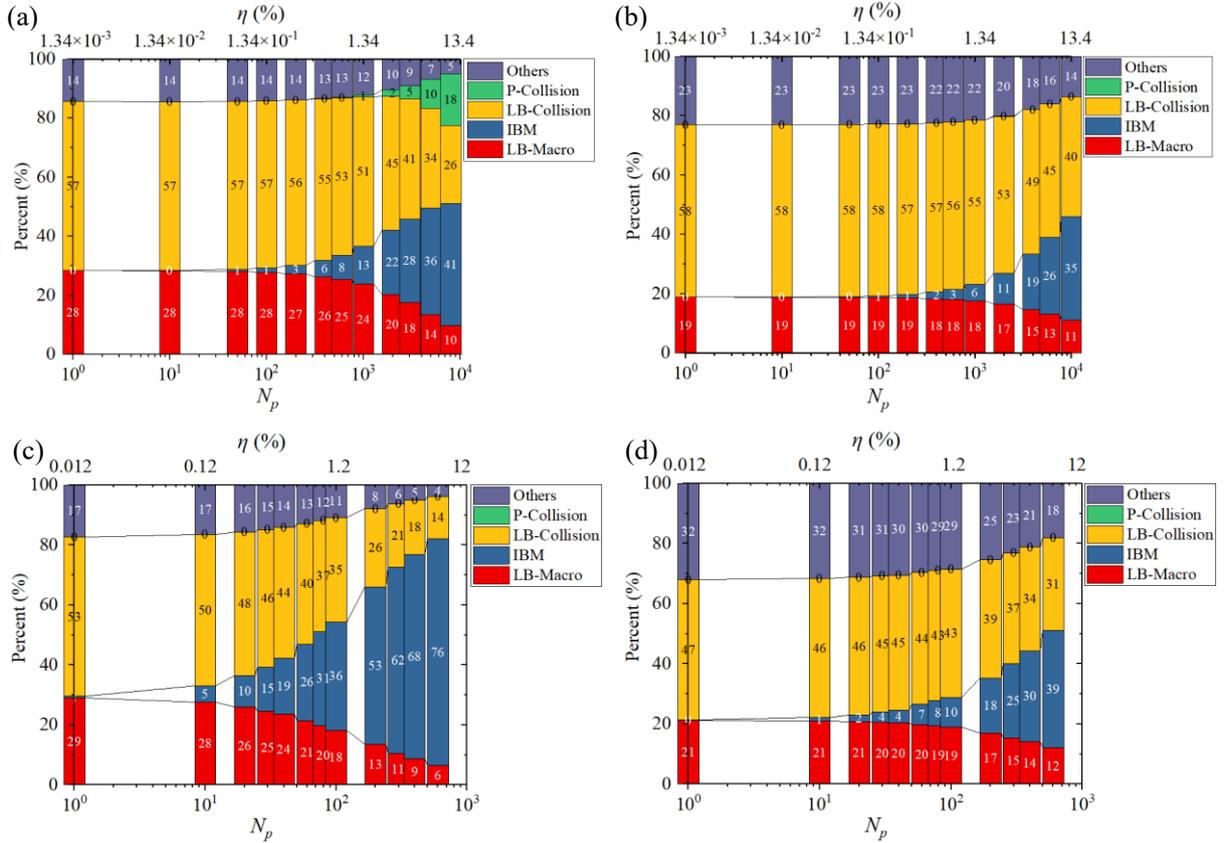

Fig. 16. Percentage of each main part of whole computing for particle-laden flows. (a) standard algorithm and (b) improved algorithm for 2D test, (c) standard algorithm and (d) improved algorithm for 3D test.

6.2.2 Effect of particle resolution

When the fluid physical domain is fixed, the particle resolution will affect the mesh ratio between particle Lagrangian points to the fluid Eulerian nodes. In this section, the fluid domains are fixed to be $40D_p \times 40D_p$ and $5D_p \times 5D_p \times 5D_p$ and the particle resolution $D_p/dh$ is accordingly varied from 12 to 48 and from 12 to 30 in 2D and 3D. It is interesting that the mesh ratio decreases as the particle resolution increases. The same as the section 6.2.1, the calculation time of the improved algorithm is also much less than that of the standard algorithm as shown in Fig. 17. The increase ratio of the curves of the real calculation time is similar to that of the theoretical deduced value in Fig. 5.

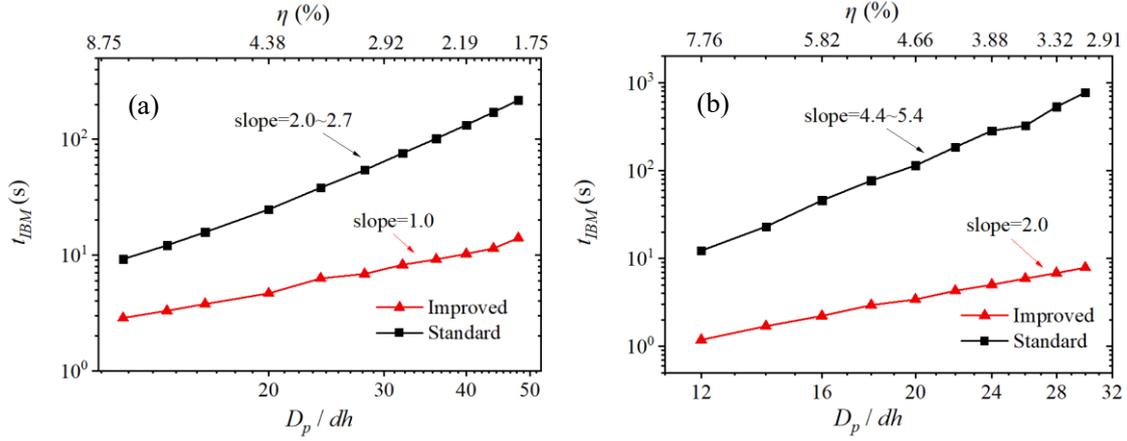

Fig. 17. Effect of particle resolution on IB calculation. (a) 2D test and (b) 3D test.

For the LB collision calculation, the slops of the curves are the same for the improved algorithm and the standard algorithm, for example 1.8~1.9 and 2.8 for 2D and 3D flows as shown in Fig. 18. This is because that for a certain mesh ratio, only the calculation amount on each fluid node are reduced while the total number of fluid nodes with IB force are not changed.

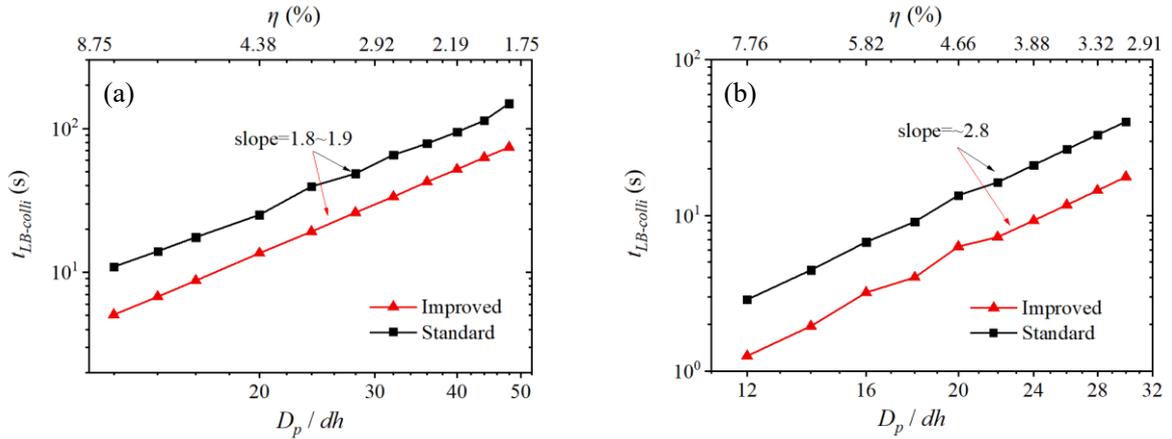

Fig. 18. Effect of particle resolution on LB collision calculation. (a) 2D test and (b) 3D test.

Figure 19 shows the total time consumption of the whole computing of particle-laden flows. Focusing on the slop of the curves, both the curves in Fig. 19(a) are similar to those in Fig.18(a). However, the curve obtained by the standard algorithm is similar to that in Fig. 17(b) and the curve obtained by the improved algorithm is similar to that in Fig.18(b). This can be explained as follows. For the two-dimensional simulations at this range of mesh ratio, the dominate calculation is LB collision for both algorithms. While for the three-dimensional simulations at this range of mesh ratio, the dominate calculations are the IB calculation and the LB collision for the standard algorithm and the improved algorithm, respectively.

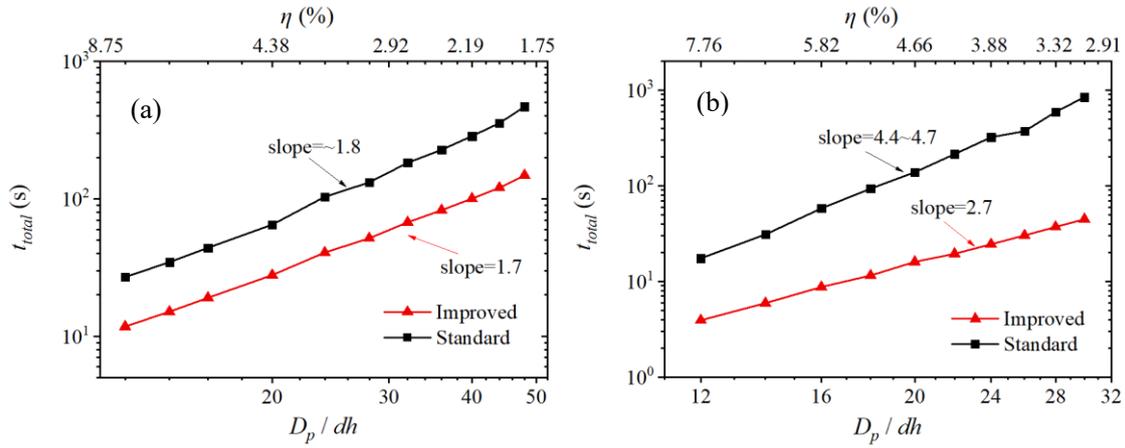

Fig. 19. Effect of particle resolution on whole calculation. (a) 2D test and (b) 3D test.

Fig. 20 shows the percentage of each main part calculation with different particle resolutions at the low mesh ratio ($\eta<10\%$). It indicates that the time consumption of IB part of standard algorithm increases with the particle resolution increasing even the mesh ratio decreasing. The IB part dominates the whole calculation, especially for the 3D flows as can be seen in Fig. 20(c). However, for the improved algorithm the IB part decreases as the particle resolution increases either in 2D or 3D. For the low mesh ratio, the LB collision part always dominates the whole calculation.

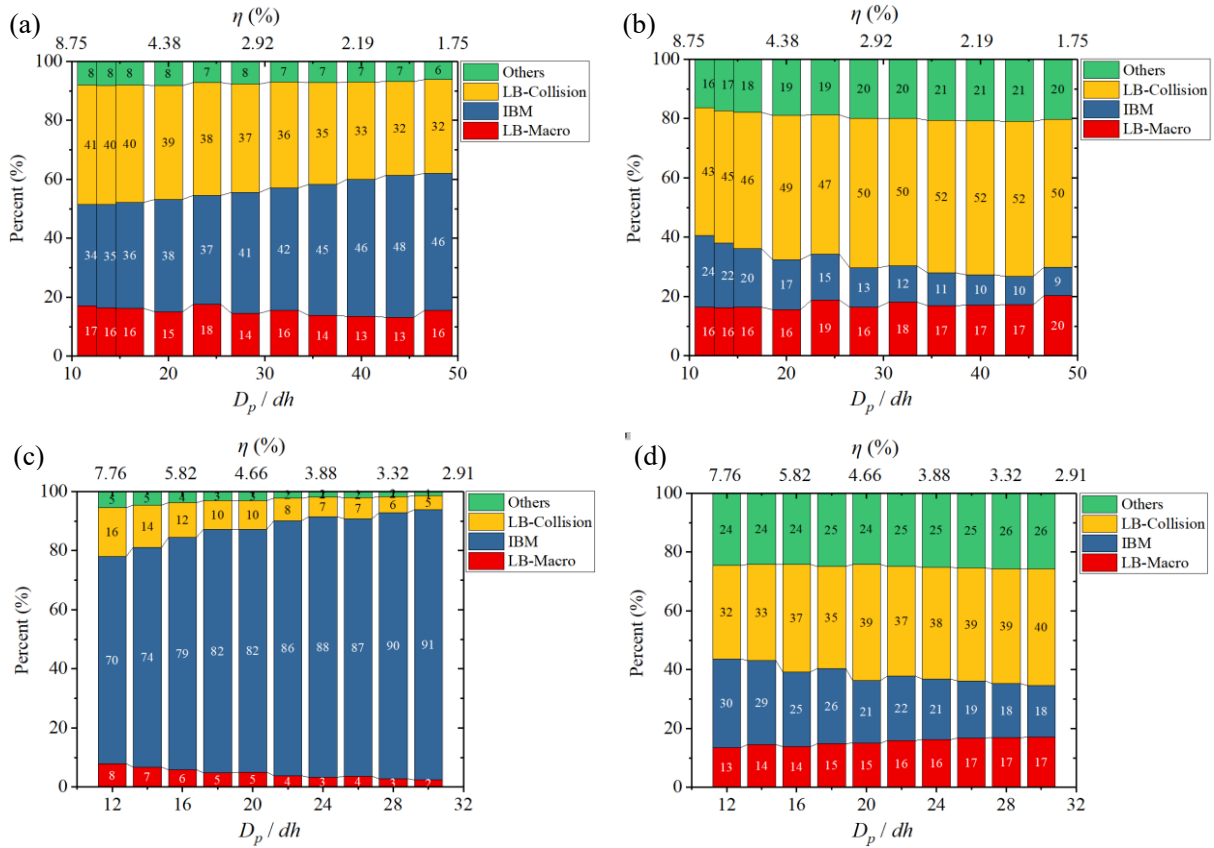

Fig. 20. Percentage of each main part of whole computing for particle-laden flows. (a) standard algorithm and (b) improved algorithm for 2D test, (c) standard algorithm and (d) improved algorithm for 3D test.

## 7. Conclusion and discussions

We have demonstrated the importance of considering the single-CPU performance before using parallel computing methods in the framework of a CFD code based on the IB-LBM in 2D and in 3D particle-laden flows [7]. In this paper, we have explored three different aspects to optimize the immersed boundary-lattice Boltzmann method algorithm for simulation of massive particle-laden flows. We first looked into the implementation of LBM, propose unrolling technique, symmetric technique and swap algorithm to reduce the solve of the macro quantities, the equilibrium distribution function, the external force term and the streaming process with low memory storage. We then transitioned to propose a local point-to-node algorithm to minimum the calculation of immersed boundary method for the interaction between fluid and solid particles on the boundary. Third, we discussed the optimization of search strategy of particle-pair short range interaction based on the grid-search algorithm. The performance of the new improved algorithm is compared to the standard algorithm based on different scale simulations. For the particle-free flows, we have obtained MLUPS of 36 in 2D simulations and 12 in 3D simulations using the improved algorithm. The speedup factors to the standard algorithm is 1.7~2.3 and at the same time has saved nearly half memory consumption. For the particle-laden flows, the reflected relative performance efficiency of the improved algorithm increases as the particle volume fraction and the particle diameter resolution increasing. The improved algorithm can achieve MLUPS of no lower than 15 and 7 in 2D and 3D dense flows, respectively.

In summary, the proposed optimizing techniques for IB-LBM calculation are potentially good for simulations of large-scale dense flows with massive particles and the high Reynolds number flows with high particle resolution. It also good for the simulations needing multi-PDFs model, such as the chemical reaction simulation with many chemical components and the non-Newtonian flows, in which the configuration tensor should also be represented by PDFs. By the way, these optimizations also can be applicable to the TRT or MRT model. In addition, though only lubrication force model is used for the particle-pair shot range interaction, it is undoubtedly to be used for the direct contact and collision model such as soft-sphere DEM model.

In this paper, we only studied the performance based on a single CPU core. Anyway, we can find that all improving algorithms in Section 3-5 for each calculation part are extremely local. Hence, the good single process performance obtained in this Part I work is a solid foundation for the massively parallel simulation, which is our another work in Part II.

**Acknowledgments**

This work is supported by the National Natural Science Foundation of China (NSFC) (Grant Nos. 51876075, 51876076).

**Supplementary**

S1. The contour of IB force on fluid grids when 2D particle moving.

S2. The contour of IB force on fluid grids when 3D particle moving.